\documentclass[amsmath,amssymb,12pt,preprint]{revtex4}
\usepackage{graphicx}

\newcommand{\bsigma}{\mbox{\boldmath $\sigma$}}
\newcommand{\bham}{\mbox{\boldmath $H$}}
\newcommand{\state}[1]{\vert #1\rangle}

\begin{document}

\title{Relational Physics and Quantum Space}
\date{\today}
    \author{Olaf Dreyer}
    \email{odreyer@perimeterinstitute.ca}
    \affiliation{Perimeter Institute for Theoretical Physics, 35 King
    Street North, Waterloo, Ontario N2J 2W9, Canada}

\begin{abstract} In a purely relational theory there exists a tension between the relational character of the theory and the existence of quantities like distance and duration. We review this issue in the context of the Leibniz-Clarke correspondence. We then address this conflict by showing that a purely relational definition of length and time can be given, provided the dynamics of the theory is known. We further show that in such a setting it is natural to expect Lorentz transformations to describe the mapping between different observers. We then comment on how these insights can be used to make progress in the search for a theory of quantum gravity. 
\end{abstract}

\maketitle

\section{Introduction}
In our current search for a quantum theory of gravity it is widely believed that the final theory should be purely relational. A long-standing thorny issue for a relational theory is the question of how quantities like distances and duration can be defined or emerge in a purely relational manner. 

This tension first became apparent in the correspondence between Clarke and Leibniz\cite{alex}. We will review that part of the correspondence that is concerned with the nature of space and time and let it be our introduction to the problem of recovering the notions of distance and duration in a relational theory. Leibniz stated his position in \cite[Third paper, \S4]{alex}: 

\begin{quote} ``\ldots I hold space to be something merely relative, as time is; \ldots I hold it to be an order of coexistences, as time is an order of succesions.''  
\end{quote}

\noindent Using his principle of the \emph{identity of indiscernibles} Leibniz then goes on to demonstrate that an absolute view of space and time is untenable and that the relative view is the only sensible one. Clarke, not at all convinced, offers the following refutation of Leibniz's position \cite[Third reply, \S4]{alex}:

\begin{quote}
``If space was nothing but the order of things coexisting; it would follow, that if God should remove in a straight line the whole world entire, with any swiftness whatsoever; yet it would still always continue in the same place: and that nothing would receive any shock upon the most sudden stopping of that motion. And if time was nothing but the order of succession of created things; it would follow, that if God had created the world millions of ages sooner than he did, yet it would not have been created at all sooner. Further: space and time are quantities; which situations and order are not.''
\end{quote}

\noindent To a modern mind this argument given by Clarke looks rather vacuous and Leibniz's reply could be given by a physicist trained today \cite[Fourth paper, \S 13]{alex}:

\begin{quote}
``To say that God can cause the whole universe to move forward in a straight line, or in any other line, without making otherwise any alteration in it; is another chimerical supposition. For, two states indiscernible from each other, are the same state; and consequently. 'tis a change without any change. \ldots''
\end{quote}

\noindent Clarke does not acknowledge this argument. Instead he concludes that Leibniz's position is disproved \cite[Fourth reply, \S 16 and \S17]{alex}:

\begin{quote}
``That space and time are not the mere order of things, but real quantities has been proven above, and no answer yet given to those proofs. And till an answer be given to those proofs, this learned author's assertion is a contradiction.''
\end{quote}

\noindent Having held to his position for four papers Leibniz now commits two grave mistakes within the space of two pages. The first one is the admission that there is '\ldots an absolute true motion \ldots' \cite[Fifth paper, \S 53]{alex}

\begin{quote}
``\ldots However, I grant there is a difference between an absolute true motion of a body, and a mere relative change of its situation with respect to another body. \ldots''
\end{quote}

\noindent If that was not enough Leibniz goes on in the next paragraph to say that distances are fundamental \cite[Fifth paper, \S 54]{alex}

\begin{quote}
``\ldots As for the objection that space and time are quantities, or rather things endowed with quantity; and that situation and order are not so: I answer, that order also has its quantity; there is in it, that which goes before and that which follows; there is distance or interval. \ldots''
\end{quote}

\noindent All Clarke has to do now is to collect his trophy. With the magnanimity of the victor he points out \cite[Fifth reply, \S 53]{alex}

\begin{quote}
``Whether this learned author's being forced here to acknowledge the difference between absolute real motion and relative motion, does not necessarily infer that space is really a quite different thing from the situation or order of bodies; I leave to the judgement of those who shall be pleased to compare what this learned writer here alleges, with what Sir Isaac Newton has said in the Principia, \ldots''
\end{quote}

\noindent Somewhat more triumphantly he continues in the next paragraph \cite[Fifth reply, \S 54]{alex}:

\begin{quote}
``I had alleged that time and space were \textsc{quantities}, which situation and order are not. To this, it is replied; that \emph{order has its quantity; there is that which goes before, and that which follows; there is distance and interval}. I answer: going before, and following, constitutes situation or order: but the distance, interval, or quantity of time or space, wherein one thing follows another, is entirely a distinct thing from the situation or order, and does not constitute any quantity of situation or order: the situation or order may be the same, when the quantity of time or space intervening is very different.''
\end{quote}

\noindent Thus ends the correspondence between Leibniz and Clarke with a clear defeat for the relativists. If one looks at the arguments that have been presented it is not so much a defeat but more of a self-destruction of Leibniz.

In this article we will propose to resolve the tension between quantity and relation using a simple model from solid state physics. In section \ref{sec:relation} we use a background-independent formulation of the Heisenberg spin chain as a simple model of the universe. In section \ref{sec:poincare} we show that observers \emph{inside} the system can use the excitations of the model (without reference to a lattice spacing) to define distances purely relationally. We also show that the maps between observers are naturally given by Poincar\'e transformations. This leads us to interpret this model as a ``quantum Minkowski space''. We briefly discuss consequences of this argument for the problem of quantum gravity, as well as certain observations about the precise relationship of our model to Minkowski space in the Conclusions. 

\section{A Relational Solid State Model}\label{sec:relation}
How else could the Leibniz-Clarke correspondence have gone? How could the tension between quantity and relation have been resolved? How is one to obtain the notion of distance in a purely relational manner?

The first thing to realize is that the tension between quantity and relation can not be resolved by relying on kinematics alone. Given a dynamical degree of freedom like a  traveling mode one can use it to define the notion of distance by defining how much it travels in a certain amount of time, i.e. by defining its velocity. This is just how we define the unit of length today, namely by setting the speed of light (see \cite{nist}). What is needed is a distinctive set of traveling degrees of freedom or excitations, that can be used to define the notion of distance in this way. 

To be concrete we will look at a particular model, the Heisenberg spin chain and its higher dimensional generalizations. Its Hamiltonian is given by
\begin{equation}\label{eqn:hamil}
  \bham = \sum_{(i j)} \bsigma_i\cdot\bsigma_j,
\end{equation}
where $\bsigma = (\sigma_x, \sigma_y, \sigma_z)$, and the $\sigma$'s are the Pauli matrices. For nearest neighbor interactions the lowest lying excitations are of the form
\begin{equation}\label{eqn:ansatz}
  \state{\psi} = \sum_{n} a_n \state{0\cdots 1\cdots 0},
\end{equation}
where the $0$'s and $1$'s denote eigenvectors of $\sigma_z$ and the $1$ occurs at the $n$-th position. The $a_n$'s take the form
\begin{equation}\label{eqn:as}
  a_n = e^{i\delta_k n}, 
\end{equation}
and 
\begin{equation}
	\delta_k = 2\pi \frac{k}{N},\ \ \ k=1,\ldots,N,
\end{equation}
where $N$ is the number of lattice sites. Solving the Schr\"odinger equation gives the eigenvalues of $\bham$ as a function of $k$:
\begin{equation}\label{eqn:energyk}
  E = -NA + 4 A \left( 1 - \cos \left(2\pi \frac{k}{N}\right)\right).
\end{equation}
We will denote the corresponding eigenvectors by $\vert k\rangle$. 

This is a set of traveling degrees of freedom of the model. In the next section we will use these to give a purely relational definition of distance. Note that the excitations we have introduced are perfectly well-defined \emph{without the introduction of a lattice spacing}.

\section{Poincar\'e Transformations} \label{sec:poincare}
Using the excitations of the model described above we can now proceed to define quantities like distance in a completely relational manner. This can be done by picking one particular excitation and assigning a speed to it. A length is then defined to be the amount the excitation has travelled in a certain time interval.  

A distinctive excitation in our model is given by the fastest wave packets. These excitations are of the form
\begin{equation}
\sum_k f(k)\,\vert k\rangle,
\end{equation}
where $f(k)$ is peaked around that value of $k$ for which $dE/dk$ is maximal. These wave packets are distinguished by the fact that no other excitation can overtake them. All observers, which can also be thought of as excitations of the system, will agree on that, independent of the way they themselves move. This characterization is thus completely relational.

Since observers in the spin model have only the above excitations at their disposal to explore their world there is no way for them to tell whether they are moving or resting with respect to the lattice. It is thus consistent and natural for all of them to assign the \emph{same} speed to these excitations. 

\begin{figure}
  \begin{center}
  \includegraphics[height=6cm]{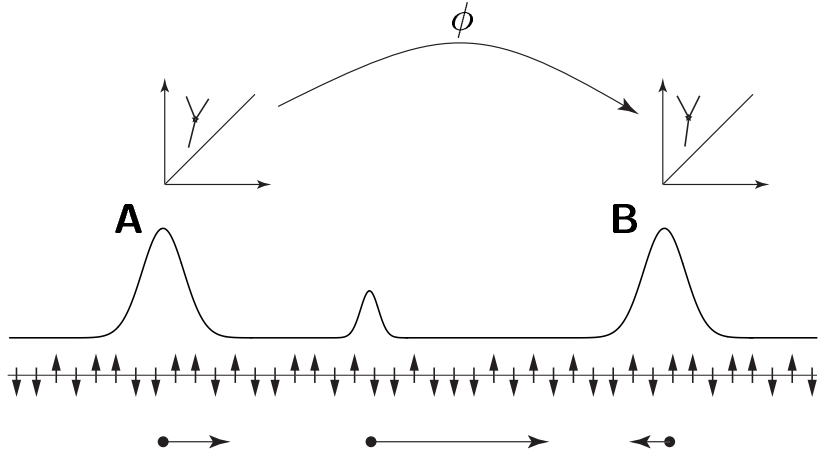}
  \end{center}
  \caption{ A view of the system that is not available to
  observers confined inside the system. The observers
  \textbf{\textsf{A}} and \textbf{\textsf{B}}, here represented by
  the large Gaussian excitations, have no way of telling what
  their motion is with respect to the lattice. This is why it is
  consistent for both observers to assign the \emph{same} speed to
  the excitation. There exists a map $\phi$ between the two
  coordinate systems given by the mapping of physical events onto each other.
  This map $\phi$ will have the property that it maps the fastest 
  excitations onto fastest excitations. We find then that this
  map $\phi$ must be a Poincar\'e transformation. \label{fig:relativ}}
\end{figure}

What will be the map between the coordinate systems of two observers? In the limit that the spin system looks smooth to the observers we can answer this question. The map can be constructed by mapping physical events onto each other. This map will in particular map the fastest excitations in one coordinate system onto the fastest excitations in the other system. Since these excitations have the same speed in both systems the map can only be a Poincar\'e transformation (see figure \ref{fig:relativ}). One thus obtains a ``quantum Minkowski space''. 

\section{Conclusion}
As it was pointed out by Clarke in his correspondence with Leibniz, in a completely relational theory there exists a tension between quantities and relations. We have seen here how this tension can be resolved provided one has access to excitations that can be used to define the notion of distance by defining the speed of these excitations. A consequence of this definition is that the natural mapping between two observers is given by a Poincar\'e transformation.

To define a notion of distance in a relational way it was necessary to have access to the dynamics of the theory. A purely kinematic approach is not sufficient. It is here where some of the candidate theories of quantum gravity, like Loop Quantum Gravity\cite{thie} or Causal Set Theory\cite{causal}, face their greatest problems. The arguments presented here suggest that the dynamics of the theory is required to make progress on important issues like the question of the semi-classical limit of the theory.

We conclude by remarking that the model presented above does deviate from usual Minkowski space in two ways. In order for us to find Poincar\'e transformations we assumed that the excitations involved are all well separated from each other. If this is not the case an operational definition of distance and duration can not be given anymore. Another deviation occurs when the observers have access to excitations with very high values of $k$ (these excitations are not to be confused with the fastest excitations used above). In this case the observers would notice the spin lattice and would find measurable deviations from Poincar\'e invariance.

\end{document}